\documentstyle[12pt,epsfig]{article}
\newcommand{\z}{&\hspace*{-8pt}}
\newcommand{\eps}{\varepsilon}
\newcommand{\e}{\varepsilon}
\newcommand{\G}{\Gamma}

\begin{document}

\begin{center}
{\Large \bf Special case of sunset: reduction and $\eps$-expansion.}
\\ \vspace*{5mm} A.~Onishchenko\footnote{on leave of absence from
Institute for High Energy Physics (Protvino, Russia) and
Institute for The\-o\-re\-ti\-c\-al and Experimental 
Physics (Moscow, Russia)}, 
O.~Veretin 
\end{center}

\begin{center}
Institut f\"ur Theoretische Teilchenphysik, \\
Universit\"at Karlsruhe, D-76128 Karlsruhe, Germany
\end{center}

\vspace*{1cm}

\abstract{
We consider two loop sunset diagrams with two
mass scales $m$ and $M$ at the threshold and pseudotreshold
that cannot be treated by earlier published formula.
The complete reduction to master integrals
is given. The master integrals are evaluated as series in
ratio $m/M$ and in $\eps$ with the help of differential equation method.
The rules of asymptotic expansion in the case when $q^2$ is
at the (pseudo)threshold are given.}\\

\section{Introduction}

  The sunset diagram plays a key role in the 2-loop calculations
with masses. Despite the fact that a lot of investigation has been
devoted to the sunset diagram there still remain drawbacks.
In Ref.~\cite{Tarasovgeneralized} general reduction procedure
is given in the case, when external momentum $q$ and internal masses are
arbitrary. But in the case when $q^2$ is equal to the threshold
or one of the pseudothresholds values the formula of 
\cite{Tarasovgeneralized} become unapplicable. 
Therefore we turn to paper \cite{DavydSmirn} where the reduction
was given specifically for the (pseudo)threshold kinematics.
However in two cases shown in Fig.1 the reduction of \cite{DavydSmirn}
fails.

  In present paper we consider calculation of these two special cases.
These integrals naturally arise in the threshold problems with 
given mass hierarchy. An immediate typical example, 
where it is the case, is a problem of matching vector 
and axial QCD currents to NRQCD ones with two heavy quark
mass scales  $m\ll M$. Another example is the calculation of 
masses of the heavy
gauge bosons in 2-loop approximation \cite{Jegerlehner:2001fb}, 
when $q^2$ is equal to $m_Z^2$ or $m_W^2$.

  For the calculation of master integrals there are mainly three
methods: 1) direct evaluation using $\alpha$ or Feynman parameter
representation, 2) solving master differential equation in
external Mandelstam variables, which can be written for the master
integrals of any Feynman graph, 3) applying various asymptotic
expansions \cite{Smirngen,book1,Smirnov:pj}. Here, we will demonstrate the
strong and weak features  of the each mentioned method on the
example of our particular problem and will advocate that a certain
mixing of these methods can give us a desired answer in the
most easiest way.

  Now let us introduce notation, which will be used later in this paper
and define master integrals we are going to calculate. For
the two-loop sunset with arbitrary masses and propagator indices
we have
\begin{equation}
J_{\nu_1\nu_2\nu_3}(q^2) = 
  \frac{1}{\pi^d} \int\!\!\!\int 
    \frac{d^dk \,\, d^dl}{[k^2-m_1^2]^{\nu_1}[(k-l)^2-m_2^2]^{\nu_2}
        [(l-q)^2-m_3^2]^{\nu_3}} \,,
\end{equation}
where $d=4-2\varepsilon$ is the dimension of space-time.

  We will discuss only two special cases of the sunset
integrals which are shown in Fig.1:
\begin{enumerate}
\item $m_1=M$, $m_2=0$, $m_3=m$, $q^2=(M+m)^2$ (threshold case),
\item $m_1=m$, $m_2=m$, $m_3=M$, $q^2=M^2$ (pseudothreshold case).
\end{enumerate}
\begin{figure}[ht]
\centerline{\vbox{\hbox{\epsfysize=30mm \epsfbox{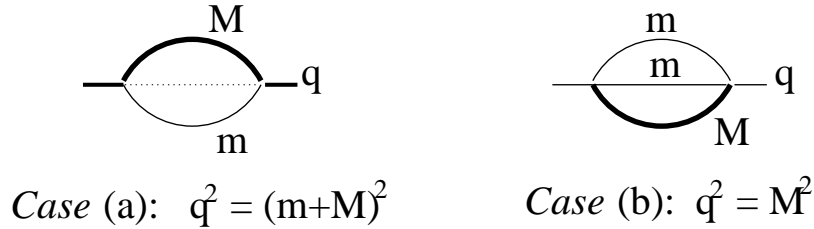}}}}
\caption{Sunset diagrams}
\end{figure}

 With the help of recurrence relations \cite{IBP}, which will be given in
next section, any integrals of the first type with arbitrary
propagator indices can be reduced to two master integrals:
$J_{111}((m+M)^2)$ and $J_{112}((m+M)^2)$, whereas in
the second case to two master integrals: $J_{111}(M^2)$ and $J_{211}(M^2)$.

  Another problem is the evaluation of the master integrals themselves.
The general representation for the sunset diagram was obtained in 
Ref.~\cite{Lauricella} in terms of hypergeometric Lauricella function. 
For the practical purposes
however one needs the $\epsilon$-expansion of these formula which
is not a trivial task. 

The result for the threshold and pseudothreshold
values of the sunset integrals with three arbitrary masses
has been ontained in Ref.~\cite{Ussykina} up to $O(1)$ 
in $\eps$-expansion. From these results we can easily obtain
the finite parts of our integrals. However we established that
in the reduction procedure for the matching of QCD to NRQCD 
currents one has to know also $O(\eps)$ part
for $J^{(a)}$ integrals and $O(\eps^2)$ part for $J^{(b)}$ integrals.
It is not easy to push forward the approach of \cite{Ussykina}
in order to evaluate these needed parts. Instead we use the 
differential equation method \cite{Kotikov}-\cite{Remiddi}
(see also \cite{Czyz}). 
We solve differential equations as series in $m/M$ and 
the desired number of
coefficients can be always obtained.
In order to find the boundary conditions for the solutions
one can use two methods: the representation of 
Ref.~\cite{Lauricella} can
be expanded in the limit $m/M\to0$ and the asymptotic expansion
procedure can be applied at the threshold and pseudothreshold.


\section{Recurrence relations at the threshold and pseudothreshold}

Here we give the recurrence relations \cite{IBP} for the sunsets of 
two types introduced earlier. 
While the derivation of these recurrence relations is straightforward,
they were not considered in the literature in detail till now.
On the other hand
they represent missing pieces to complete the generalized recurrence
relations of Tarasov \cite{Tarasovgeneralized} and threshold relations of 
Davydychev and Smirnov \cite{DavydSmirn}
for the sunset diagrams in the case of threshold with one zero mass
and pseudothreshold with two equal masses. Most of the formula below
can be derived just by combining and reexpressing 
appropriate recurrence relations of 
\cite{Tarasovgeneralized} and \cite{DavydSmirn}.

\vspace*{0.2cm}

\subsection{JM0m}
In this case we can not directly apply recurrence 
relations for general masses obtained in \cite{DavydSmirn}.
The point is as one of mass becomes zero the relations of \cite{DavydSmirn}
are degenerate. Instead we derive:
\begin{equation}
\label{jM0mrec1}
(d-2\nu_2-2) \, \nu_2 {\bf 2^+} = 
     - 2m^2 \nu_3(\nu_3+1)\, {\bf 3^{++}} 
     + (d-2\nu_3-2) \, \nu_3 {\bf 3^+}   \,,
\end{equation}
\begin{equation}
\label{jM0mrec2}
2 M^2 \nu_1(\nu_1+1) \, {\bf 1^{++}} = 
      2m^2 \nu_3(\nu_3+1)\, {\bf 3^{++}} 
     - (d-2\nu_3-2) \, \nu_3 {\bf 3^+}
     + (d-2\nu_1-2) \, \nu_1 {\bf 1^+}   \,,
\end{equation}
\begin{eqnarray}
\label{jM0mrec3}
\z\z
2 m^2 (M+m) (3d-2\nu_1-2\nu_3-7) \nu_3(\nu_3+1) {\bf 3^{++}} = 
\nonumber\\
\z\z\qquad
  \;\;\; \Bigl[ m \Bigr( (d-\nu_3-3)(2d-\nu_1-2\nu_3-4)
       + (d-\nu_1-\nu_3-2)(3d-2\nu_2-2\nu_3-6) \Bigl)
\nonumber\\
\z\z\qquad
       + M (2d-\nu_1-2\nu_3-5)(2d-\nu_1-2\nu_3-4) \Bigr] \,\nu_3 {\bf 3^+} 
\nonumber\\
\z\z\qquad
      + (d-\nu_3-2) \Bigr( m(d-\nu_3-3) 
          + M(2d-\nu_1-2\nu_3-5) \Bigr) \, \nu_1{\bf 1^+}
\nonumber\\
\z\z\qquad
      + 2 m^2 M \nu_1 \nu_3 (\nu_3+1) {\bf 1^+2^-3^{++}}
\nonumber\\
\z\z\qquad
      + \Bigl[ -m(d-\nu_3-3) - M(2d-\nu_1-2\nu_3-5) \Bigr]
         \, \nu_1\nu_3 {\bf 1^+2^-3^+}    \,,
\end{eqnarray}
\begin{eqnarray}
\label{jM0mrec4}
\z\z
2mM(m+M)(3d-11) \, J_{212} = 
 \nonumber\\
\z\z\qquad
      -(d-3) \Bigl( m(d-3) + M(2d-7) \Bigr) \, J_{112}
      -(d-3) \Bigl( m(2d-7) + M(d-3) \Bigr) \, J_{211}
\nonumber\\
\z\z\qquad
   + \frac{m+M}{4m^2M^2}(d-2)^2(2d-7) \, J_{101}   \,.
\end{eqnarray}
  Here as usual $\bf j^+$ (or $\bf j^-$) means the operator raising
(or diminishing) index on the $j$-th line.

  With the help of (\ref{jM0mrec1})-(\ref{jM0mrec4}) we reduce
all integrals to three sunset diagrams $J_{111},\,J_{112}$ and $J_{211}$
plus products of 1-loop tadpoles.
In addition there is a relation between these three sunset 
integrals and one of them can be eliminated
\begin{equation}
M(m+2M) \, J_{211} + m(M+2m) \, J_{112} = 
    \frac{3d-8}{2} \, J_{111} + \frac{(d-2)^2}{4mM(d-3)} \, J_{101} \,.
\end{equation}
This finishes the reduction procedure.


\subsection{JmmM}

Here we are faced with pseudothreshold problem and again in this case 
general recurrence relations by Tarasov \cite{Tarasovgeneralized}
and those for the threshold problems by
Smirnov and Davydichev \cite{DavydSmirn} become degenerate.
The reduction procedure for this type of integrals has been
studied in \cite{DavGro}, where the topology with three and four
lines has been considered. These integrals  were also considered
in \cite{AvdeevKalmy}, where asymptotic expansion has been used.
For the completeness we give alternative reduction formula below.

  First, using relations
\begin{eqnarray}
\label{jmmMrec1}
\z\z
2m^2 \nu_1(\nu_1+1) \, {\bf 1^{++}}  = 
     (d-2\nu_1-2) \, \nu_1 {\bf 1^+}
   - (d-2\nu_3-2) \, \nu_3 {\bf 3^+}
   + 2 M^2 \nu_3 (\nu_3+1) \, {\bf 3^{++}}  \,,\\
\z\z
2m^2 \nu_2(\nu_2+1) \, {\bf 2^{++}}  = 
     (d-2\nu_2-2) \, \nu_2 {\bf 2^+}
   - (d-2\nu_3-2) \, \nu_3 {\bf 3^+}
   + 2 M^2 \nu_3 (\nu_3+1) \, {\bf 3^{++}}  \,,
\end{eqnarray}
we reduce indices of lines 1 and 2 to one or two. Thus there
remain only integrals $J_{11\nu_3},\, J_{12\nu_3}$ and $J_{22\nu_3}$.

  For $J_{12\nu_3}$ we have
\begin{equation}
4(M^2-m^2) \, \nu_3 {\bf 3^+}  = 
    \Bigl[ {\bf 1^-} - (d-3) \, {\bf 2^-} \Bigr] \, \nu_3 {\bf 3^+} 
    - {\bf 1^+ 3^-} + d-3\nu_3   \,.
\end{equation}

  For $J_{22\nu_3}$ we have
\begin{eqnarray}
\z\z
4 m^2 (M^2-m^2) (d-\nu_3-3) = 
\nonumber\\
\z\z\qquad\qquad
     \;\;\; m^2 (d-\nu_3-3) \, {\bf 3^-} 
   + \Bigl[ m^2 \Bigl( -3d^2 + d(13+7\nu_3) 
\nonumber\\
\z\z\qquad\qquad
     - 12 - 17\nu_3 - 4\nu_3^2 \Bigr)
          + M^2 (2d-\nu_3-6)(d-\nu_3-2) \Bigr] \, {\bf 1^-} 
\nonumber\\
\z\z\qquad\qquad
     + \Bigl[ -m^2 + M^2 (d-\nu_3-2) \Bigr] (d-3)\, \nu_3 {\bf 3^+1^-2^-}
\nonumber\\
\z\z\qquad\qquad
     + \Bigl[ m^2(d-3) - M^2 (2d-\nu_3-6) \Bigr] \, \nu_3 {\bf 3^+1^{--}} \,.
\end{eqnarray}

  For $J_{11\nu_3}$ we have
\begin{eqnarray}
\z\z
4 M^2 (M^2-m^2) (d-\nu_3-3) \nu_3 (\nu_3+1) \, {\bf 3^{++}} = 
\nonumber\\
\z\z\qquad\qquad
     \;\;\; \Bigr[ - m^2 (d-2\nu_3-3)(2d-2\nu_3-5) 
\nonumber\\
\z\z\qquad\qquad
           + M^2 \Bigl( 3d^2 - d(17+7\nu_3) + 24 + 19\nu_3 + 4\nu_3^2 \Bigr)
         \Bigr] \, \nu_3 {\bf 3^+}
\nonumber\\
\z\z\qquad\qquad
     + \Bigr[ - m^2 \Bigl( d^2 - d(3+5\nu_3) + \nu_3(4\nu_3+11) \Bigr)
       - M^2 (d-\nu_3-2) \nu_3 \Bigr] \, {\bf 2^+}
\nonumber\\
\z\z\qquad\qquad
    + \Bigr[ -m^2(d-3) + M^2 \nu_3 \Bigr] \, \nu_3 {\bf 3^+2^+1^-}
    + m^2 (d-\nu_3-3) \, {\bf 3^-1^+2^+}  \,.
\end{eqnarray}

  Finaly there is relation between three integrals
\begin{eqnarray}
M^2 J_{112} + m^2 J_{211} = 
   \frac{3d-8}{4} J_{111}
          - \frac{(d-2)^2}{8 m^2 (d-3)} J_{110}  \,.
\end{eqnarray}
  This finishes the reduction.


\section{Master differential equation}

It is known, that a general two-loop sunset topology has four master
integrals \cite{Tarasovgeneralized}: 
one with unit powers of propagators and three with a dot
placed on one of the lines. For master integrals considered
in this paper this number is however smaller, which is due to a symmetry of
the mass distribution on the lines. In addition in case of $JM0m$
the index of massless line can be always reduced 
to 1 (see \cite{Tarasovgeneralized}).
It can be shown, that these 
master integrals satisfy a sytem of linear nonhomogeneous 
differential equations in $q^2$ ($q$ being external momentum) \cite{Remiddi}
or in masses \cite{Kotikov}. 
However, in our case we want to write differential equations in 
$r=m/M$, where $m$ and $M$ are internal masses and
$q^2$ lays at threshold or pseudothreshold. Differential equations of 
this type were considered earlier in \cite{FraTausk}.

\subsection{JM0m}

In this case we have two master integrals ($J_{111}$ and $J_{112}$)
and the system of two 1st order equations. It is convinient to 
rescale integrals introducing $\tilde{J}_{111}$ and $\tilde{J}_{112}$ by
\begin{eqnarray}
\label{tildeJ}
J_{111} = M^{2d-6}\Gamma^2(3-d/2)\, \tilde{J}_{111}\quad\mbox{and}\quad
J_{112} = M^{2d-8}\Gamma^2(3-d/2)\, \tilde{J}_{112}. 
\end{eqnarray}
Then the differential equations read
\begin{eqnarray}
\left\{ 
  \begin{array}{l} 
    \displaystyle
    (r+1)(r+2)\frac{\partial}{\partial r} \tilde{J}_{111} 
       - 6r(r+1) \tilde{J}_{112} - \Bigr( d+2(d-3)r-4 \Bigl) \tilde{J}_{111} 
       - \frac{8r^{d-3}}{(d-4)^2(d-3)}  = 0 \,,\\ 
    \label{JM0msystem} \\
     \displaystyle
    (r+1)(r+2)\frac{\partial}{\partial r} \tilde{J}_{112} + 
    \frac{1}{r}(2r^2-5dr+20r-4d+14) \tilde{J}_{112} \\
     \displaystyle  \qquad\qquad\qquad \hfill
       - \frac{(d-3)(3d-8)}{2r} \tilde{J}_{111} 
       - \frac{8r^{d-5}}{(d-4)^2} = 0 \,.
  \end{array}
\right.
\end{eqnarray}
  Using this system one can write the second order 
differential equation for $\tilde{J}_{111}$: 
\begin{eqnarray}
\label{difurJ111}
\z\z (r+1)^2(r+2)^2\frac{\partial^2}{\partial r^2} \tilde{J}_{111} - 
     \frac{2}{r}(r+1)(r+2)^2 \Bigl( d+(d-4)r-3 \Bigr)
      \frac{\partial}{\partial r} \tilde{J}_{111}
\nonumber \\ 
\z\z\qquad 
  + \frac{1}{r}(d-3)^2(r+2)^2(d-2r-4) \tilde{J}_{111} - 
  \frac{8}{(d-4)^2}(r+2)^2r^{d-4} = 0. \label{JM0mequation}
\end{eqnarray} 
  As usual we search the solution of (\ref{difurJ111}) as a linear
combination of two solutions of the homogeneous equation and the solution
of the nonhomogeneous equation. We will find the solution as a
series in $r$, namely, in the following form
\begin{eqnarray}
\tilde{J}_{111} = \sum_i r^{\alpha_i}\left(\sum_{n=0}^{\infty}
    a_n^{(i)}(d) \,r^n\right) . \label{JM0mexpand}
\end{eqnarray}

  By substituting (\ref{JM0mexpand}) in Eq. (\ref{JM0mequation}), we find
for leading exponents $\alpha_i$ three allowed values $\alpha_1 = 0$,
$\alpha_2 = 2d-5$, corresponding to the two independent solutions
of the associated homogeneous equation and $\alpha_3 = d-2$, as required
by the nonhomogeneous part of Eq. (\ref{JM0mequation}). 
So the only thing we need yet to do is to find
the coeffients in front of two independent solutions
of homogeneous part of Eq. (\ref{JM0mequation}) using boundary conditions. 
One equation for
these coefficients can be obtained from the value of $J_{111}$ master
integral at $r = 0$, which can be written in 
terms of $\Gamma$-functions.
One can {\it not} use $J_{112}$ at $r = 0$ as the second boundary condition.
The reason is that the latter integral becomes
infrared divergent and this divergency is regularized by $\epsilon$ 
($d = 4-2\epsilon$) and not by $r$ as in the limit $r\to0$.
In order to obtain a second boundary condition we need explicit
expansion of function $J_{111}$ in $r$ up to the third order, which can be
obtained by analysing representation of this integral in terms of
Appel function $F_4$ or performing asymptotic expansion in small mass
ratio $r$. Expansion of the Appel function will be considered in Appendix A,
while asymptotic expansion technique is discussed in Section 4.1.

Having obtained expansion in $r$ up to order $O(r^3)$
using one of the mentioned methods, one can easily reconstruct all
other expansion coefficients from differential equation.
Returning back to functions $J_{111}$ and $J_{112}$
instead of $\tilde{J}_{111}$ and $\tilde{J}_{112}$
we give the first seven coefficients of expansion:
\begin{eqnarray}
\label{JM0m111result}
\z\z e^{2\epsilon\gamma_E}M^{-2+4\epsilon}J_{111} = 
-\frac{1+r^2}{2\epsilon^2} + \frac{(8L-5)r^2+2r-5}{4\epsilon} -
\frac{1}{8}(11+20\zeta_2) + \frac{5r}{4} \nonumber\\
\z\z -\frac{1}{8}(16L^2-48L-12\zeta_2+7)r^2 - 
\frac{2}{9}(9L^2+15L+54\zeta_2-8)r^3 + 
\frac{1}{8}(24L^2+20L+144\zeta_2-3)r^4 \nonumber\\
\z\z-\frac{2}{225}(450L^2+255L+2700\zeta_2-16)r^5
+\frac{1}{72}(360L^2+156L+2160\zeta_2-5)r^6 \nonumber\\
\z\z+\epsilon\left( 
\frac{55}{16}-\frac{25\zeta_2}{4}-\frac{11\zeta_3}{3} + 
\frac{1}{8}(20\zeta_2+11)r + \left( 
\frac{4L^3}{3}-6L^2+2(\zeta_2+7)L\right.\right.\nonumber\\
\z\z\left.+\frac{1}{48}(276\zeta_2+208\zeta_3
+321)\right)r^2 + \left(4L^3+\frac{2L^2}{3}+(40\zeta_2-\frac{238}{9})L
-\frac{68\zeta_2}{3}-24\zeta_3+\frac{773}{54}\right)r^3\nonumber\\
\z\z+\left(-6L^3-\frac{L^2}{2}+(\frac{463}{12}-60\zeta_2)L
+17\zeta_2+36\zeta_3-\frac{4639}{144}\right)r^4\nonumber\\
\z\z+\left(8L^3+\frac{23L^2}{15}+(80\zeta_2-\frac{22681}{450})L
-\frac{134\zeta_2}{15}-48\zeta_3+\frac{1242497}{27000}\right)r^5 
\nonumber\\\z\z\left.+\left(-10L^3-\frac{19L^2}{6}
+(\frac{3751}{60}-100\zeta_2)L-\frac{5\zeta_2}{3}
+60\zeta_3-\frac{648161}{10800}\right)r^6\right) + O(r^7) + O(\eps^2)
\end{eqnarray}
and
\begin{eqnarray}
\z\z e^{2\epsilon\gamma_E}M^{4\epsilon}J_{112} = -\frac{1}{2\epsilon^2}
+\frac{2L-\frac{1}{2}}{\epsilon}+\frac{1}{4}(6\zeta_2+2)+4L-2L^2 
+(-2L^2-4L-12\zeta_2+2)r\nonumber\\
\z\z+(3L^2+3L+18\zeta_2-\frac{1}{2})r^2
+(-4L^2-\frac{8L}{3}-24\zeta_2+\frac{2}{9})r^3
+(5L^2+\frac{5L}{2}+30\zeta_2-\frac{1}{8})r^4\nonumber\\
\z\z+(-6L^2-\frac{12L}{5}-36\zeta_2+\frac{2}{25})r^5
+(7L^2+\frac{7L}{3}+\frac{1}{18}(756\zeta_2-1))r^6\nonumber\\
\z\z+\epsilon\left(\frac{1}{12}(66\zeta_2+52\zeta_3+66)
+8L+2L\zeta_2-4L^2+\frac{4L^3}{3}
+(4L^3+4L^2+40\zeta_2L-24L\right.\nonumber\\
\z\z-8\zeta_2-24\zeta_3+14)r
+(-6L^3-5L^2-60\zeta_2L+37L-6\zeta_2+36\zeta_3-\frac{67}{2})r^2 
\nonumber\\
\z\z+\left(8L^3+\frac{22L^2}{3}+80\zeta_2L-\frac{440L}{9}
+\frac{68\zeta_2}{3}-48\zeta_3+\frac{1277}{27}\right)r^3\nonumber\\
\z\z+\left(-10L^3-\frac{31L^2}{3}-100\zeta_2L+\frac{1097L}{18}
-42\zeta_2+60\zeta_3-\frac{4403}{72}\right)r^4\nonumber\\
\z\z+\left(12L^3+\frac{69L^2}{5}+120\zeta_2L-\frac{5489L}{75}
+\frac{318\zeta_2}{5}-72\zeta_3+\frac{75179}{1000}\right)r^5
\nonumber \\
\z\z+\left.\left(-14L^3-\frac{529L^2}{30}-140\zeta_2L+\frac{6418L}{75}
-\frac{1307\zeta_2}{15}+84\zeta_3-\frac{4823273}{54000}\right)r^6\right)
+ O(r^7) + O(\eps^2), \nonumber\\
\end{eqnarray}
where $L=\log r$. 
The $O(1)$ part of $J_{111}$ is in agreement with \cite{Ussykina}.


\subsection{JmmM}

This diagram was recently studied in Ref. \cite{Argeri} by means of
differential equation method in the regime when $m\gg M$, while
we are interested in the case $m\ll M$.
Here we also have two master integrals and the system of two
differential equations. Again introducing rescaled functions
according to (\ref{tildeJ}) we have
\begin{eqnarray}
\left\{
 \begin{array}{l}
  \displaystyle
  \frac{\partial}{\partial r} \tilde{J}_{111} - 4r \tilde{J}_{211} = 0 \,,\\
   \label{JmmMsystem} \\
   \displaystyle
  (r^2-1)\frac{\partial}{\partial r} \tilde{J}_{211}
  +\frac{r^2(13-4d)+2d-7}{r} \tilde{J}_{211}
  +\frac{(d-3)(3d-8)}{4r} \tilde{J}_{111} 
  -\frac{2r^{d-7}(r^d-2r^2)}{(d-4)^2} = 0 .
 \end{array}
\right.
\end{eqnarray}

Using (\ref{JmmMsystem}) the corresponding second
order differential equation for $J_{111}$ looks like
\begin{eqnarray}
(r^2-1)\frac{\partial^2}{\partial r^2}J_{111}
-\frac{2(d-3)(2r^2-1)}{r}\frac{\partial}{\partial r}J_{111}
+(d-3)(3d-8)J_{111} - \frac{8r^{d-6}(r^d-2r^2)}{(d-4)^2} =0. 
\label{JmmMequation}
\end{eqnarray}
To find a solution of this equation we again use Ansatz 
(\ref{JM0mexpand}) for the most general form of solution at $r\to0$. 
For leading exponents $\alpha_i$ there are four allowed values 
$\alpha_1 = 0$, $\alpha_2 = 2d-5$, corresponding to the two independent
solutions of the associated homogeneous equation and 
$\alpha_3 = d-2$, $\alpha_4 = 2d-4$, as required by nonhomogeneous
part of Eq. (\ref{JmmMequation}). All other steps in this case are in 
one to one correspondence with those considered in previous subsection.
The first boundary condition
is given by the value of master integral $J_{111}$ at $r=0$. 
In order to find the second boundary condition we need explicit expression for
$r$-expansion of $J_{111}$ up to the third order, which can
be obtained from threshold large mass  assymptotic expansion of $J_{111}$. 
As a result of all these steps we have the following expressions for
our master integrals
\begin{eqnarray}
\z\z e^{2\epsilon\gamma_E}M^{-2+4\epsilon}J_{111} = 
-\frac{1+2r^2}{2\epsilon^2}+\frac{4(4L-3)r^2-5}{4\epsilon}
-\frac{1}{8}(11+20\zeta_2)-(4L^2-12L-3\zeta_2+5)r^2 \nonumber \\
\z\z-\frac{1}{4}(8L^2-12L+8\zeta_2+7)r^4 - \frac{1}{18}(12L-11)r^6
+\epsilon\left(\frac{55}{16}-\frac{11\zeta_3}{3}-\frac{25\zeta_2}{4}
\right. \nonumber\\
\z\z+\left(\frac{8L^3}{3}-12L^2+4(\zeta_2+7)L+9\zeta_2 
+\frac{26\zeta_3}{3}-3\right)r^2 - 32\zeta_2r^3 \nonumber \\
\z\z\left.+\left(4L^3-5L^2-\frac{7L}{2}+4\zeta_2-4\zeta_3 
+\frac{95}{8}\right)r^4+\frac{32\zeta_2r^5}{5}
+\frac{2}{9}(8L-9\zeta_2-14)r^6\right)\nonumber\\
\z\z+\epsilon^2\left(\frac{949}{32}-\frac{55\zeta_2}{8}-\frac{55\zeta_3}{6}
-\frac{1}{720}\pi^4(296r^4-578r^2+303)\right. \nonumber \\
\z\z+\left(-\frac{4L^4}{3}+8L^3-4(\zeta_2+7)L^2
+\frac{4}{3}(9\zeta_2-2\zeta_3+45)L+31\zeta_2+26\zeta_3+19\right)r^2
\nonumber \\
\z\z+\frac{64}{3}(6L+12\log 2-11)\zeta_2r^3
+\left(-\frac{14L^4}{3}+6L^3+(\frac{3}{2}-2\zeta_2)L^2
+(3\zeta_2+\frac{37}{4})L\right. \nonumber\\
\z\z\left.-\frac{71\zeta_2}{4}+8\zeta_3-\frac{885}{16}
\right)r^4 - \frac{32}{75}(60L+120\log 2-77)\zeta_2r^5 \nonumber\\
\z\z\left.+\frac{1}{324}(288L^3-792L^2-108(2\zeta_2+11)L+1926\zeta_2
-1296\zeta_3+5417)r^6\right) + O(r^7) + O(\eps^3)
\end{eqnarray}
and
\begin{eqnarray}
\z\z e^{2\epsilon\gamma_E}M^{4\epsilon}J_{211} = 
-\frac{1}{2\epsilon^2}+\frac{4L-1}{2\epsilon}+\frac{1}{2}
+\frac{3\zeta_2}{2}+4L-2L^2-(2L^2-2L+2\zeta_2+1)r^2\nonumber \\
\z\z+(\frac{3}{4}-L)r^4+(\frac{5}{36}-\frac{L}{3})r^6 
+\epsilon\left(\frac{11}{2}+\frac{11\zeta_2}{2}+\frac{13\zeta_3}{3} 
+8L+2L\zeta_2-4L^2+\frac{4L^3}{3}-24\zeta_2r\right.\nonumber \\
\z\z+(4L^3-2L^2-6L+4\zeta_2-4\zeta_3+11)r^2 + 8\zeta_2r^3
+\frac{1}{9}(24L-27\zeta_2-38)r^4 + \frac{8\zeta_2r^5}{5}
\nonumber \\
\z\z\left.+(\frac{L}{45}-\zeta_2-\frac{83}{300})r^6\right)
+\epsilon^2\left(\frac{49}{2}+\frac{37\zeta_2}{2}+\frac{37\zeta_3}{3}
+\frac{289\pi^4}{720}-\frac{4}{3}L(\zeta_3-12)+4L\zeta_2 
\right.\nonumber \\
\z\z-2L^2\zeta_2-8L^2+\frac{8L^3}{3}-\frac{2L^4}{3}
+48\zeta_2(2L+4\log 2-3)r \nonumber\\
\z\z+\left(-\frac{14L^4}{3}+\frac{4L^3}{3}-2(\zeta_2-3)L^2+2(\zeta_2+5)L
-17\zeta_2+8\zeta_3-\frac{37\pi^4}{90}-53\right)r^2\nonumber \\
\z\z-\frac{8}{3}(12L+24\log 2-13)\zeta_2r^3
+\left(\frac{4L^3}{3}-3L^2-(\zeta_2+\frac{121}{18})L
+\frac{35\zeta_2}{4}-6\zeta_3+\frac{5219}{216}\right)r^4 \nonumber \\
\z\z-\frac{4}{75}(120L+240\log 2-199)\zeta_2r^5 \nonumber\\
\z\z\left.+\left(\frac{4L^3}{9}-\frac{5L^2}{9}
-\frac{1}{450}(150\zeta_2+203)L+\frac{37\zeta_2}{180}-2\zeta_3
+\frac{237511}{81000}\right)r^6\right) + O(r^7) + O(\eps^3) \,,
\end{eqnarray}
where $L=\log r$.
The $O(1)$ part of $J_{111}$ is in agreement with \cite{Ussykina}.


\section{Asymptotic large mass expansion at the threshold}

In this section we consider asymtotic large mass expansion for
master integrals introduced above. A type of expansion one needs to
perform in order to obtain an analytic expression for the master integral
of case (b) was already considered in Ref.~\cite{CzarnSmirn} and
\cite{AvdeevKalmy} and one
may just follow along the lines of procedure described there. However, 
in the case of master integral JM0m a somewhat different prescription
for setting loop momenta is required\footnote{Asymptotic
expansions in momentum space are not invariant under the redefinition
of loop momenta contrary to the expansions, performed 
in $\alpha$-representation
for Feynman integrals. Therefore special care should be taken 
choosing a correct one set of momenta.} 
and thus we will consider this case in detail below.   

In order to establish the expansion procedure we use ''the strategy
of regions'' \cite{Smirnov:1998vk}. 
Let us remind you first a general prescription for large mass expansion
at the threshold of Ref.~\cite{CzarnSmirn}. We consider a general case of 
threshold Feynman integral
$F_{\Gamma}$, corresponding to a graph $\Gamma$ when the masses $M_i$ and
external momenta $Q_i$ are considered large with respect to small masses
$m_i$ and external momenta $q_i$. We are interested in a case, when
external momenta $Q_i$ are on the following mass shell: 
$Q_i^2 = (\sum_j a_{ij}M_j + \sum_k b_{ik}m_k)^2$. Here  $a_{ij}$ and 
$b_{ik}$ are some numbers. It is just the generalization of the 
on-mass-shell condition of Ref.~\cite{CzarnSmirn}. 
Then the asymptotic expansion in the limit $Q_i, M_i\to\infty$ takes 
the following explicit form \cite{Smirngen}
\begin{eqnarray}
F_{\Gamma}(Q_i, M_i, q_i, m_i;\eps) 
\stackrel{M_i\to\infty}{\sim}
\sum_{\gamma}{\cal M}_{\gamma}F_{\Gamma}(Q_i, M_i, q_i, m_i;\eps). 
\label{generalasympt}
\end{eqnarray}
Here the sum runs over subgraphs $\gamma$ of $\Gamma$ such that
\begin{enumerate}
\item in $\gamma$ there is a path between any pair of external vertices
associated with the large external momenta $Q_i$;
\item $\gamma$ containes all the lines with the large masses;
\item  every connectivity component $\gamma_j$ of the graph $\hat\gamma$
obtained from $\gamma$ by collapsing all the external vertices with
the large external momenta to apoint is 1PI with respect to the lines
with small masses.  
\end{enumerate}

Operator ${\cal M}_{\gamma}$ in (\ref{generalasympt}) 
can be written as a product 
$\Pi_i{\cal M}_{\gamma_i}$ over different connectivity components, where 
${\cal M}_{\gamma_i}$ are operators of Taylor expansion in certain
momenta and masses. In what follows we will distinguish the connectivity 
component $\gamma_0$, which is defined to contain all external vertices with 
large momenta. For connectivity components $\gamma_i$, different 
from $\gamma_0$, the corresponding operator ${\cal M}_{\gamma_i}$ performs 
Taylor expansion of the Feynman integral $F_{\gamma_i}$ in its small masses
and external momenta. To describe the action of ${\cal M}_{\gamma_0}$ 
one uses representation of $\gamma_0$-component in terms of a union 
of its 1PI components and cut heavy lines (that is subgraph becomes 
disconnected after a cut line is removed). Here we can again factorize
${\cal M}_{\gamma_0}$ and the Taylor expansion of the 1PI components
of $\gamma_0$ is performed as in a case of other connectivity components
$\gamma_i$. As for the action of operator ${\cal M}$ on the cut lines
there are two different cases. Let $P + k$ be the momentum of a line
with large mass $M_i$, where $P$ is a linear combination of large
external momenta and $k$ is a linear combination of loop momenta
and small external momenta. Then the mentioned two cases can be written as
follows:
\begin{itemize}
\item $P^2=M_i^2$ and ${\cal M}$ for this line is given by
\begin{eqnarray}
{\cal M} = \left.{\cal T}_x\frac{1}{xk^2+2Pk}\right|_{x=1},
\end{eqnarray}
Here ${\cal T}_x$ denotes the operator of Taylor expansion in $x$ 
around $x=0$. 
\item $P^2 \neq M_i^2$ and the operator ${\cal M}$ reduces to the ordinary
Taylor expansion in small (with respect to this line) external momenta.
\end{itemize} 
As for the optimal set of internal loop momenta we choose a rule when a
large external momenta are divided between lines with masses $M_i$ and $m_k$
in order to satisfy the following conditions: $P_i^2=M_i^2$ and $P_k^2=m_k^2$.
We do not know whether such separation is always possible or not, but
in most cases of interest it certainly works. As an example in the next
subsection we will consider an expansion for master integral
$J_{111}((m+M)^2)$ from our case (a).

\subsection{Second boundary condition for JM0m}
For master integral $J_{111}$, according to our prescription for
choosing internal momenta we have the following expression:
\begin{eqnarray}
J_{111} = \frac{1}{\pi^d}\int\!\!\!\int
\frac{d^dkd^dl}{[k^2][(k+l)^2+2a(k+l,q)][l^2+2b\,lq]},
\end{eqnarray}
Here $q$ is an external momentum; $a=M/(M+m)$ and $b=-m/(M+m)$. 
From  the general formula (\ref{generalasympt}) in the asymptotic
expansion of master integral $J_{111}$ in the limit
$m/M\to 0$ we have four subgraphs\footnote{Here $\Gamma$ is the graph,
corresponding to Feynman integral $J_{111}$.}:
\begin{enumerate}
\item[1)] graph $\Gamma$ itself;
\item[2)] subgraph $\gamma_1$ consisting from lines with masses $M$ and
$m$;
\item[3)] subgraph $\gamma_2$ consisting from lines with mases $M$ and
zero;
\item[4)] subgraph $\gamma_3$ consisting from one heavy line.
\end{enumerate}

  For 1) we expand the integrand around $m=0$ with the result
\begin{eqnarray}
\sum_{n=0}^{\infty}\frac{(-1)^n(2b)^n}{\pi^d}
\int\!\!\!\int
 \frac{d^dkd^dl(l\cdot q)^n}{[k^2][(k+l)^2+2a(k+l,q)][l^2]^{n+1}}\,.
\end{eqnarray}
Each term in this expansion can be evaluated rewriting it
via scalar integrals with shifted space-time 
dimension \cite{Tarasovgeneralized} and 
for first three of them we have:
\begin{eqnarray}
\z\z\frac{1}{\pi^d}\int\!\!\!\int
\frac{d^dkd^dl(lq)}{[k^2][l^2]^2[(k+l)^2+2a(k+l,q)]} = 
-aq^2\mbox{JV}^{d+2}_{222}, \nonumber \\
\z\z\frac{1}{\pi^d}\int\!\!\!\int
\frac{d^dkd^dl(lq)^2}{[k^2][l^2]^3[(k+l)^2+2a(k+l,q)]} =
-\frac{q^2}{2}\mbox{JV}^{d+2}_{231} - \frac{q^2}{2}\mbox{JV}^{d+2}_{132}
+4a^2q^4\mbox{JV}^{d+4}_{333}, \nonumber \\
\z\z\frac{1}{\pi^d}\int\!\!\!\int
\frac{d^dkd^dl(lq)^3}{[k^2][l^2]^4[(k+l)^2+2a(k+l,q)]} = 
-36a^3q^6\mbox{JV}^{d+6}_{444} + 3aq^4\mbox{JV}^{d+4}_{243}
+3aq^4\mbox{JV}^{d+4}_{342}, \nonumber
\end{eqnarray}
where 
$$\mbox{JV}^d_{\nu_1\nu_2\nu_3} = \frac{1}{\pi^d}\int\!\!\!\int
\frac{d^dkd^dl}{[k^2]^{\nu_1}[l^2]^{\nu_2}[(k+l)^2+2(k+l,q)]^{\nu_3}}
\quad \mbox{and} \quad q^2 = M^2\,.$$
The latter can be easily expressed in terms of $\Gamma$-functions for arbitrary
values of $\nu_1$, $\nu_2$ and $\nu_3$.

The Taylor expansion for subgraph $\gamma_2$ gives 
\begin{eqnarray}
\sum_{n=0}^{\infty}\frac{(-1)^n}{\pi^d}\int\!\!\!\int
\frac{d^dkd^dl(k^2+2kl+2akq)^n}
{[k^2][l^2+2a lq]^{n+1}[l^2+2b lq]} = 0.
\end{eqnarray}
In case of subgraph $\gamma_3$ we have the following expression
\begin{eqnarray}
\sum_{n=0}^{\infty}\frac{(-1)^n}{\pi^d}\int\!\!\!\int
\frac{d^dkd^dl(l^2+2kl+2alq)^n}{[k^2][k^2+2akq]^{n+1}[l^2+2blq]}.
\end{eqnarray}
Each term of this sum is just a product of two one-loop integrals and 
hence can be easily evaluated.

An expansion for subgraph $\gamma_4$ leads to the following result
\begin{eqnarray}
\sum_{n=0}^{\infty}\frac{(-1)^n}{\pi^d}\int\!\!\!\int
\frac{d^dkd^dl((k+l)^2)^n}{[k^2][l^2+2b\,lq][2a(k+l,q)]^{n+1}}.
\end{eqnarray}
Here we can see, that  it is the most difficult from computation point of 
view type of contribution, as we are dealing in this case with eikonal
integrals. As one can easily see, the evaluation of each term from this 
contribution can be reduced to the evaluation of integrals of the
following type
\begin{eqnarray}
\frac{1}{\pi^d}\int\!\!\!\int
\frac{d^dkd^dl(kl)^m}{[k^2][l^2+2blq][(k+l,q)]^n}. \label{klsubgraph4}
\end{eqnarray}

  In order to calculate these integrals  it is naturally  to express
the products $(kl)^m$ in terms of traceless products 
$(kl)^{(m)}\equiv k^{(\alpha,m)}l_{(\alpha,m)}$\footnote{
Here $\alpha$ is a collective index representing $m$ Lorenz indices
$\alpha_1,\alpha_2,\dots,\alpha_m$}. Then we notice, that
integration over $k$ of traceless products $k^{(\alpha, m)}$ times
the part of integrand, which depends only on $k$ results in expression
propotional to the traceless product $q^{(\alpha, m)}$. Thus, we can replace
the factor $(kl)^{(m)}$ by $(qk)^{(m)}(ql)^{(m)}/(qq)^{(m)}$ and finally
replace the factors involved through ordinary products $qk$ and $lq$.
After performing these steps we reduce the evaluation of 
integrals (\ref{klsubgraph4}) to the integrals:
\begin{eqnarray}
\frac{1}{\pi^d}\int\!\!\!\int
\frac{d^dkd^dl(lq)^m}{[k^2][l^2+2b\,lq][(k+l,q)]^n}.
\end{eqnarray}
These integrals with the use of integration by parts identities 
$(d-3-n)n^+ + (n+1)m^+n^{++} = 0$, where $n(m)^+$ are operators, which
increase corresponding indices $n$ or $m$) can be further reduced to 
integrals of the form
\begin{eqnarray}
\frac{1}{\pi^d}\int\!\!\!\int
\frac{d^dkd^dl(lq)^m}{[k^2][l^2+2b\,lq][(k+l,q)]}.
\end{eqnarray}
To take these last integrals one can first perform $k_0$ and $l_0$
integrations using Cauchy theorem and remained angular integrations are
trivial since there are no products $kl$ in the integrand left.

However, as our calculations showed, it is far more simple just to
calculate using asymtotic expansion this master integral up to the order
$(m/M)^3$ and then use this result as the nessesary
boundary conditions for the solution of master differential equation.
It takes then much less efforts to construct the expansion of
this master integral up to any order in $r=m/M$ using differential
equation method.


\section{Conclusion}
We considered two loop sunset diagrams with two
mass scales $m$ and $M$ at the threshold and pseudotreshold
that cannot be threated by earlier published formula
\cite{Tarasovgeneralized,DavydSmirn}.
The complete reduction to the master integrals
is given. The master integrals are evaluated as series in
ratio $m/M$ and up to needed in applications order in $\eps$ with the help 
of differential equation method.
The rules of asymptotic expansion in the case when $q^2$ is
at the (pseudo)threshold are given. 

{\bf Acknowledgements.} We would like
to thank K. Chetyrkin, A. Davydychev, V.A. Smirnov, M.Yu. Kalmykov 
and O.V. Tarasov for fruitfull discussions of the topics 
discussed in this paper and valuable comments. Authors acknowlege
Universit\"at Karlsruhe for the warm hospitality.

\appendix

\section{Expansion of $F_4$ Appel function: second boundary condition
for JM0m}

In order to obtain the second boundary integral in case of JM0m
we can use the representation of the sunset diagram in
term of Lauricella function which can be obtained from \cite{Lauricella}.
However since one mass is zero, the Lauricella function simplifies
to Appel function:
\begin{eqnarray}
\label{appelsunset}
\z\z e^{2\gamma\varepsilon} M^{-2+4\varepsilon} 
J_{111}( m,M,0;\,q^2 ) = 
  - \left(\frac{m^2}{M^2}\right)^{1-\varepsilon}
  \Gamma^2(-1+\varepsilon)\,
  F_4 \left( { 1,\, \varepsilon \atop 2-\varepsilon,\, 2-\varepsilon} 
         \,\,;\, \frac{m^2}{M^2},\, \frac{q^2}{M^2} \right) \nonumber\\
\z\z \qquad
  - \Gamma(1-\varepsilon) \Gamma(-1+\varepsilon) \Gamma(-1+2\varepsilon)\,
  F_4 \left( { -1+2\varepsilon,\, \varepsilon \atop 
         \varepsilon,\, 2-\varepsilon} 
         \,\,;\, \frac{m^2}{M^2},\, \frac{q^2}{M^2} \right) 
\end{eqnarray}
and in our case $q^2=(m+M)^2$. We need the expansion of this
integal up to the order $O(r^3)$ and $O(\eps)$. Then the rest 
coefficients can be found from the differential equation.

  In order to expand (\ref{appelsunset}) in series over $m/M$ we
use the following representation for $F_4$ 
\begin{eqnarray}
\label{bljaF4}
  F_4 \left( { a,\, b \atop c,\, c'} 
         \,\,;\, x,\, y \right)  = 
   \sum_{k=0}^\infty \frac{(a)_k (b)_k}{(c)_k} \frac{x^k}{k!}
  {}_2F_1 \left( { a+k,\, b+k \atop c'} 
         \,\,;\, y \right) .
\end{eqnarray}
In (\ref{bljaF4}) function ${}_2F_1$ has to be transformed
to argument $1-y$. Then we have rot the r.h.s. ($r=m/M$)
\begin{eqnarray}
\label{huisunset}
\z\z
-r^{2-2\e} \G^2(-1+\e) \sum\limits_{k=0}^\infty
  \frac{(1)_k (\e)_k}{(2-\e)_k} \frac{r^{2k}}{k!} \Biggl\{ \nonumber\\
\z\z\qquad\qquad
 \G\left( {\scriptstyle {2-\e,\,1-2\e-2k \atop 1-\e-k,\,2-2\e-k}} \right)\,
 {}_2F_1 \left( {\scriptstyle { 1+k,\,\e+k \atop 2\e+2k}}; 
    -r(2+r)\right) \nonumber\\
\z\z\qquad\qquad
 + \left[ -r(2+r) \right]^{1-2\e-2k}
   \G\left( {\scriptstyle {2-\e,\,2\e+2k \atop 1+k,\,\e+k}} \right) \,
 {}_2F_1 \left( {\scriptstyle { 1-\e-k,\,2-2\e-k \atop 2-2\e-2k}}; 
        -r(2+r)\right) \Biggr\}
\nonumber\\
\z\z
- \G(-1+\e)\G(1-\e)\G(-1+2\e) \sum\limits_{k=0}^\infty
  (-1+2\e)_k \, \frac{r^{2k}}{k!} \Biggl\{ \nonumber\\
\z\z\qquad\qquad
 \G\left( {\scriptstyle { 2-\e,\,3-4\e-2k \atop 3-3\e-k,\,2-2\e-k} } \right)\,
 {}_2F_1 \left( {\scriptstyle { -1+2\e+k,\,\e+k \atop 2\e+2k}}; 
       -r(2+r)\right) \nonumber\\
\z\z\qquad\qquad
 + \left[ -r(2+r) \right]^{3-4\e-2k}
 \G\left( {\scriptstyle {2-\e,\,-3+4\e+2k \atop -1+2\e+k,\,\e+k}}\right)
 {}_2F_1 \left( {\scriptstyle { 3-3\e-k,\,2-2\e-k \atop 4-4\e-2k}}; 
        -r(2+r)\right) \Biggr\} \,.
\end{eqnarray}
 The first and third terms in (\ref{huisunset}) can be easily
expanded since the series in $k$ and ${}_2F_1$ series can be
truncated at the given order.
In the second and fourth terms we still can truncate ${}_2F_1$ series, 
however, we cannot truncate the sum over $k$ because of 
factor $[-r(2+r)]^{-2k}$. Thus we have to resum the whole $k$-sum.
Thus in order $O(r^3)$ we have for the second term
$$
 \frac{(-r)^{-2\e} (+r)^{-2\e}}{2\sqrt{\pi}} 
 \G(2-\e)\G^2(-1+\e)\G(-1/2+\e)\,  
 {}_2F_1 \left( {\scriptstyle { 3-3\e-k,\,2-2\e-k \atop 4-4\e-2k}}; 
        1\right)
$$
and the fourth term
$$
 \frac{(-r)^{-4\e}}{2\sqrt{\pi}\G(\e)} 
 \G(2-\e)\G^2(1-\e)\G(-1+\e)\G(-3/2+2\e)\G(-1+2\e)\,  
 {}_2F_1 \left( {\scriptstyle { 3-3\e-k,\,2-2\e-k \atop 4-4\e-2k}}; 
        1\right) .
$$
Each of this two terms has an imaginary part but it cancels in the
sum of the two. Adding contributions from the first and third terms of
(\ref{huisunset}) and expanding in $\e$ we get $O(r^3)$ term
of the formula (\ref{JM0m111result}).


\end{document}